\documentclass[graybox]{svmult}

\usepackage{mathptmx}       
\usepackage{helvet}         
\usepackage{courier}        
\usepackage{type1cm}        
\usepackage{makeidx}         
\usepackage{graphicx}        
\usepackage{multicol}        
\usepackage[bottom]{footmisc}

\makeindex             

\begin{document}

\title*{Imaging transverse electron focusing in semiconducting heterostructures with
spin-orbit coupling}
\titlerunning{Imaging transverse electron focusing in ...}
\authorrunning{Reynoso et al.}

\author{A. A. Reynoso, Gonzalo Usaj and C. A. Balseiro}
\institute{A. A. Reynoso \at Instituto Balseiro and Centro At\'omico Bariloche,
Comisi\'on Nacional de Energ\'\i a At\'omica, 8400 San Carlos de Bariloche,
Argentina, \email{reynoso@cab.cnea.gov.ar}. \and Gonzalo Usaj \at Instituto
Balseiro and Centro At\'omico Bariloche, Comisi\'on Nacional de Energ\'\i a
At\'omica, 8400 San Carlos de Bariloche, Argentina. \and C. A. Balseiro \at
Instituto Balseiro and Centro At\'omico Bariloche, Comisi\'on Nacional de
Energ\'\i a At\'omica, 8400 San Carlos de Bariloche, Argentina.}
%
%
\maketitle

\abstract*{Transverse electron focusing in two-dimensional electron gases
(2DEGs) with strong spin-orbit coupling is revisited. The transverse focusing is
related to the transmission between two contacts at the edge of a 2DEG when a
perpendicular magnetic field is applied. Scanning probe microscopy imaging
techniques can be used to study the electron flow in these systems. Using
numerical techniques we simulate the images that could be obtained in such
experiments. We show that hybrid edge states can be imaged and that the outgoing
flux can be polarized if the microscope tip probe is placed in specific
positions. }

\abstract{Transverse electron focusing in two-dimensional electron gases (2DEGs)
with strong spin-orbit coupling is revisited. The transverse focusing is related
to the transmission between two contacts at the edge of a 2DEG when a
perpendicular magnetic field is applied. Scanning probe microscopy imaging
techniques can be used to study the electron flow in these systems. Using
numerical techniques we simulate the images that could be obtained in such
experiments. We show that hybrid edge states can be imaged and that the outgoing
flux can be polarized if the microscope tip probe is placed in specific
positions.}

\section{Introduction}
\label{sec:1} During the last decade, a tremendous amount of work has been
devoted to manipulate and control the spin degree of freedom of the charge
carriers \cite{Spintronicsbook}. It was quickly recognized that the spin-orbit
(SO) interaction may be a useful tool to achieve this goal. This is due to the
fact that the SO coupling links currents, spins and external fields. Using
intrinsic material properties to control the carrier's spin would allow one to
build spintronic devices without the complication of integrating different
materials in the same circuit \cite{Spintronicsbook}. The challenging task of
building spin devices based purely on semiconducting technology, requires one to
inject, control and detect spin polarized currents. During the last years a
number of theoretical and experimental papers were devoted to the study of the
effect of SO coupling on the electronic, magnetic and magnetotransport
properties of 2DEGs (see \cite{ReynosoUB06} and references therein). The nature
of the SO coupling in these systems is due to the Dresselhauss and the Rashba
mechanisms, the latter being the dominant effect in several cases
\cite{Winkler_book}. In addition, the Rashba coupling has the advantage that its
strength can be changed when a gate voltage is applied to the heterostructure,
opening new alternatives for device design \cite{NittaATE97}.

In many transport experiments in 2DEG with a transverse magnetic field,
including quantum Hall effect and transverse magnetic focusing, the SO coupling
plays a central role. The transverse focusing consists basically
in injecting carriers at the edge of a 2DEG and collecting them at a distance $%
L $ from the injection point. The propagation from the injector $I$ to the
detector $D$ is ballistic and the carriers can be focalized onto the detector by
means of an external magnetic field perpendicular to the 2DEG. The field
dependence of the focusing signal is essentially given by the transmission from
$I$ to $D$. In a semiclassical picture, the trajectories that dominate the
focusing signal are semicircles whose radius can be tuned with the external
field. The new scanning technologies developed in
\cite{TopinkaLSHWMG00,TopinkaLWSFHMG01} can be used to map these trajectories.
The scanning probe imaging techniques consist in perturbing the system with the
tip of a scanning microscope and plotting the transmission as a function of the
tip position. The transmission change is a map of the electron flow. In this
paper we first revisit the theory of transverse electron focusing in systems
with strong SO coupling and interpret the results in terms of a simple
semiclassical picture \cite{UsajB04_focusing}. Then, we use numerical techniques
to simulate the images that could be obtained with scanning probe microscopy
experiments. We show that hybrid edge states can be visualized and that the
outgoing flux can be polarized if the microscope tip probe is placed in specific
positions

\section{Transverse electron focusing in presence of strong spin-orbit
coupling} \label{sec:2}

The Hamiltonian of a 2DEG with Rashba spin-orbit coupling is given by
\begin{equation}
H\!=\!\frac{1}{2m^{\ast }}(P_{x}^{2}\!+\!P_{y}^{2})\!+\!\frac{\alpha }{\hbar
}(P_{y}\sigma _{x}\!-\!P_{x}\sigma _{y})\!-\!\frac{1}{2}g\mu _{B}\sigma
_{z}B_{z}\!+\!V(x)  \label{HRB}
\end{equation}
here $P_{\eta }\!=\!p_{\eta }\!-\!(e/c)A_{\eta }$ with $p_{\eta }$ and $%
A_{\eta }$ being the $\eta $-component of the momentum and vector potential
respectively, $\alpha $ is the Rashba coupling parameter, $g$ is the effective
g-factor, $\{\sigma _{\eta }\}$ are the Pauli matrices and $V(x)$ describes the
potential at the edge of the sample. In what follows, we use a hard wall
potential: $V(x)\!=\!0$ for $x\geq 0$ and infinite otherwise. For convenience we
choose the vector potential in the Landau gauge $\mathbf{A}\!=\!(0,xB_{z},0)$.

Far from the sample edge ($x\!\gg\!0$) the eigenvalues and eigenfunctions of
Hamiltonian (\ref{HRB}) are well known \cite{BychkovR84}. The SO coupling breaks
the spin degeneracy of the Landau levels. The spectrum is given by
\begin{equation}
E_{n}^{\pm }\!=\!\hbar \omega _{c}n\mp \sqrt{E_{0}^{2}+\left( \frac{\alpha }{%
l_{c}}\right) ^{2}2n}\,,  \label{ener}
\end{equation}
where $n\geq 1$, $\omega _{c}\!=\!e\left| B\right| /m^{*}c$ is the cyclotron
frequency, $l_{c}\!=\!(\hbar /m\omega _{c})^{\frac{1}{2}}$ is the magnetic
length, and $E_{0}\!=\!\hbar \omega _{c}/2-g\mu _{B}B_{z}/2$ is the energy of
the ground multiplet corresponding to $n\!=\!0$. The eigenfunctions for $n\geq
1$, written as spinors in the $z$-direction, are \cite{note1}
\begin{equation}
\Psi _{n,k}^{+}(x,y)\!=\!\frac{1}{\sqrt{A_{n}L_{y}}}e^{\mathrm{i}ky}\left(
\begin{array}{c}
\phi _{n-1}(x-x_{0}) \\
-D_{n}\phi _{n}(x-x_{0})
\end{array}
\right)
\end{equation}
and
\begin{equation}
\Psi _{n,k}^{-}(x,y)\!=\!\frac{1}{\sqrt{A_{n}L_{y}}}e^{\mathrm{i}ky}\left(
\begin{array}{c}
D_{n}\phi _{n-1}(x-x_{0}) \\
{\phi _{n}(x-x_{0})}
\end{array}
\right) \,,
\end{equation}
where $L_{y}$ is the length of the sample in the $y$-direction, $\phi
_{n}(x-x_{0})$ is the harmonic oscillator wavefunction centered at the
coordinate $x_{0}=l_{c}^{2}k$ , $A_{n}\!=\!1+D_{n}^{2}$ and

\begin{equation}
D_{n}\!=\!\frac{\left( \alpha / l_{c}\right) \sqrt{2n}}{E_{0}+\sqrt{%
E_{0}^{2}+\left( \alpha / l_{c}\right) ^{2}2n}}\;.  \label{dn}
\end{equation}

The wave functions of the first Landau level are given by $\Psi _{n,k}^{-}(x,y)$
with $n=0$.

These eigenstates have a cyclotron radius given by
\begin{equation}
r_{c}^{2}\!=\!2\left\langle \Psi _{n}^{\pm }|(x-x_{0})^{2}|\Psi _{n}^{\pm
}\right\rangle ,
\end{equation}
that for large $n$ gives $r_{c}^{2}\simeq $ $2n(\hbar /m^{\ast }\omega _{c})$%
. We see from Eq.(\ref{ener}) that states with different $n$, and consequently
different cyclotron radius, coexist within the same energy
window. Additionally, in the limit of strong Rashba coupling or large $n$, $%
D_{n}\!\sim 1$ and the spin lies in the plane of the 2DEG.

Equivalent results are found in a semiclassical treatment of the problem \cite
{PletyukhovAMB02,ReynosoUSB04}. In this approach, the spin is described by a
vector \cite{PletyukhovAMB02} $\mathbf{S}=\hbar /2(n_{1}(t),n_{2}(t),n_{3}(t))$
and the classical orbits are given by
\begin{eqnarray}
\mathbf{q} &=&r_{\pm }(\cos {\omega _{\pm }}t,\sin {\omega _{\pm }}t)
\nonumber  \label{semicl} \\
\mathbf{S} &=&\mathrm{sign}(B_{z})\frac{\hbar }{2}(\mp \cos {\omega _{\pm
}}t,\mp \sin {\omega _{\pm }}t,0),
\end{eqnarray}
here $\mathbf{q}$ is the coordinate measured from the centre of the circular
orbit of radius
\begin{equation}
r_{\pm }\!=\!\sqrt{\left( \frac{\alpha }{\hbar \omega _{c}}\right) ^{2}+%
\frac{2E}{m^{\ast }{\omega _{c}}^{2}}}\pm \frac{\alpha }{\hbar \omega _{c}}%
\;,  \label{rs}
\end{equation}
and the corresponding cyclotron frequencies are
\begin{equation}
\omega _{\pm }=\mathrm{sign}(B_{z})(\omega _{c}\mp {\alpha }/{\hbar r_{\pm
}})\;. \label{omegapm}
\end{equation}
In agreement with the quantum results obtained for large $n$, the spin is found
to be in-plane pointing outwards for the smaller orbit and inwards for the
bigger one when a positive perpendicular magnetic field $B_{z}$ is applied.
Moreover, the Born-Sommerfeld quantization \cite{GutzBook} of these periodic
orbits reproduces the exact quantum results of Eq.(\ref{ener}) for large $n$.

The calculation of the exact edge states with the hard wall potential requires a
numerical approach. We have shown that the semiclassical approximation can be
extended to describe edge states in which electrons bounce at the sample edge
\cite{ReynosoUSB04}. Due to the continuity of the wave function and
the spin conservation at the edge, the two orbits with radii $r_{+}\!$ and $%
r_{-}\!$ are mixed as schematically shown in Fig.1. The agreement between the
Born-Sommerfeld quantization of the semiclassical edge states and the quantum
results is excellent for states composed of semicircles centered in the edge
(normal incidence). In what follows, we use these semiclassical orbits to
interpret the numerical results for transverse focusing experiments.

The transverse focusing experiments collect electrons or holes coming from a
point contact \cite{PotokFMU02,RokhinsonLGPW04} into another point contact
acting as a voltage probe. The carriers are focused onto the collector by the
action of an external magnetic field as schematically shown in Fig. 1. The
signal measured in transverse focusing experiments is related to the
transmission $T$ between the two point contacts located at a distance $L$ from
each other (see Fig. 1). Typical experimental setups also include two ohmic
contacts at the bulk of the 2DEG which are used to inject currents and measure
voltages. The details of different configurations with four contacts have been
analyzed in \cite{vanHouten89}. The main features of the magnetic field
dependence of the focusing peaks are contained in $T$ \cite {BeenakkerH91}.
Consequently, from hereon we will refer to the focusing signal or to $T$
indistinctly. In the zero temperature limit we only need to evaluate $T$ at the
Fermi energy $E_F$. For the numerical calculation of $T$ the system was
discretized using a tight-binding model in which the leads or contacts are
easily attached. In this approach the Hamiltonian is given by $H=H_{0}+H_{SO}$
with
\begin{equation}
H_{0}\!=\!\sum_{n,\sigma }\varepsilon _{\sigma }c_{n\sigma }^{\dagger
}c_{n\sigma }^{}-\sum_{<n,m>,\sigma }t_{nm}\,c_{n\sigma }^{\dagger }c_{m\sigma
}^{}\!+\!h.c.\,.
\end{equation}

\begin{figure}[t]
  \begin{center}
     \includegraphics[width=.8\textwidth]{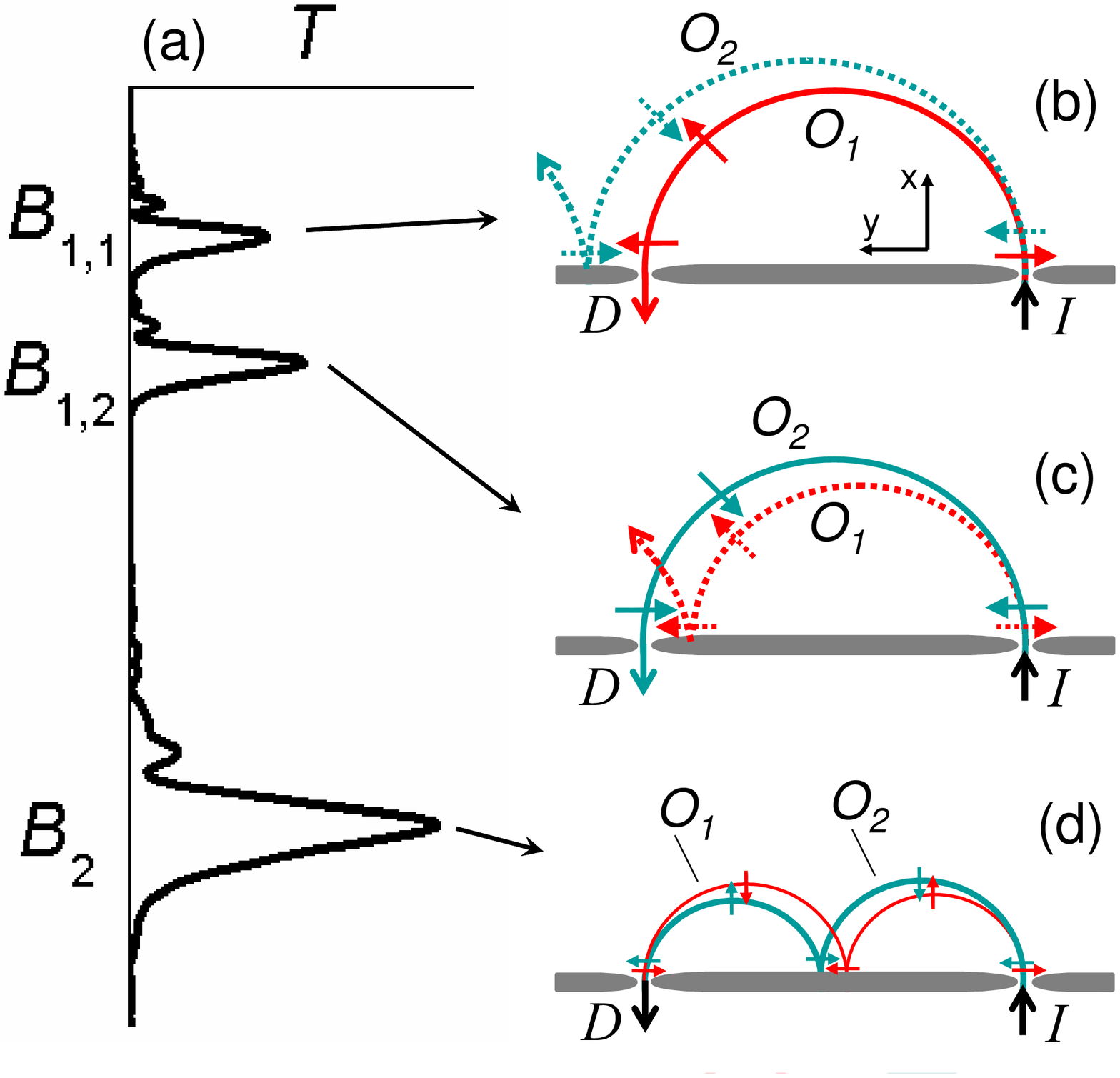}
     \caption{Panel(a) Transmission coefficient between the contacts $I$ and $D$ as a function of the applied perpendicular magnetic
     field in the presence of strong Rashba spin-orbit coupling (qualitative). Relevant semiclassical orbits for three different focusing conditions are shown in panels (b),(c) and (d).  }
   \label{Figu1}
  \end{center}
\end{figure}

Here $c_{n\sigma }^{\dagger }$ creates an electron at site $n$ with spin $%
\sigma $ ($\uparrow $ or $\downarrow $ in the $z$ direction) and energy $%
\varepsilon _{\sigma }\!=\!4t\!-\!\sigma g\mu _{B}B_{z}/2$, $t\!=\!-$ $\hbar
^{2}/2m^{*}a_{0}^{2}$ and $a_{0}$ is the lattice parameter which is always
chosen small compared to the Fermi wavelength. The summation is made on a
square lattice, the coordinate of site $n$ is $n_{x} \widehat{\mathbf{x}} + n_{y}%
\widehat{\mathbf{y}}$ where $\widehat{\mathbf{x}}$ and $\widehat{\mathbf{y}}$
are unit vectors in the $x$ and $y$ directions, respectively. The hard-wall
potential
$V(x)$ is introduced by taking $n_{x}>0$. The hopping matrix element $%
t_{nm}$ connects nearest neighbors only and includes the effect of the
diamagnetic coupling through the Peierls substitution \cite{Ferrybook}. For
the choice of the Landau gauge $t_{n(n+\widehat{\mathbf{y}})}\!=\!t\exp {(%
\mathrm{i}n_{x}2\pi \phi /\phi _{0})}$ and $t_{n(n+\widehat{\mathbf{x}})}\!=\!t$%
, $\phi \!=\!a_{0}^{2}B_{z}$ is the magnetic flux per plaquete and $\phi
_{0}\!=\!hc/e$ is the flux quantum. The second term of the Hamiltonian
describes the spin-orbit coupling, 
\begin{eqnarray}
H_{SO}&\!=\!&\sum_{n}\left\{ \lambda _{x}c_{n\uparrow }^{\dagger }c_{(n+%
\widehat{\mathbf{x}})\downarrow }-\lambda _{x}^{*}c_{n\downarrow }^{\dagger
}c_{(n+\widehat{\mathbf{x}})\uparrow }+\right.   \label{Htb} \\
&&\left. e^{\mathrm{i}n_{x}2\pi \phi /\phi _{0}}\left[ \lambda _{y}c_{n\uparrow
}^{\dagger }c_{(n+\widehat{\mathbf{y}})\downarrow }\!-\!\lambda
_{y}^{*}c_{n\downarrow }^{\dagger }c_{(n+\widehat{\mathbf{y}})\uparrow }\right]
\right\} \!+\!h.c.  \nonumber
\end{eqnarray}
where $\lambda _{x}\!=\!\alpha /2a_{0}$ and $\lambda _{y}=-\mathrm{i}%
\alpha /2a_{0}$. In what follows we use the following values for the microscopic
parameters: $a_{0}=5nm$, $m^{*}=0.055m_{0}$---here $m_{0}$ is
the free electron mass---and $E_{F}=23meV$. These parameters correspond to $%
InAs$ based heterostructures with a moderate doping. We use different values of
the SO coupling parameter $\alpha $ as indicated in each case.

The two la\-te\-ral contacts, $I$ (injector) and $D$ (detector) are attached to
the semi-infinite 2DEG described by Hamiltonian (\ref{Htb}). Each contact is an
ideal (with $\!\alpha =0$) narrow stripe of width $N_{0}a_{0}$. They represent
point contacts gated to have a single active channel with a conductance
$2e^{2}/h$,  for details see \cite{UsajB04_focusing}.

To obtain the transmission between the two contacts we calculate the Green
functions between the sites of the injector and the sites of the detector.
As the spin is not conserved, the Green function between two sites $i$ and $%
j $ has four components $\mathcal{G}_{i\sigma ,j\sigma ^{\prime }}$. First the
propagators of the system without the contacts are obtained by Fourier
transforming in the $y$-direction and generating a continuous fraction for each
$k$. Having these propagators, the self energies due to the contacts can be
easily included using the Dyson equation \cite{Ferrybook}. The transmission is
then obtained as

\begin{equation}
T=\!\frac{e^{2}}{h}\mathrm{Tr}\left[\Gamma ^{(2)}\mathcal{G}%
^{R}\Gamma ^{(1)}\mathcal{G}^{A}\right]_{\omega =E_{F}}
\end{equation}
here $\mathcal{G}^{R}$ and $\mathcal{G}^{A}$are the retarded and advanced Green
function matrices with elements $\mathcal{G}_{i\sigma ,j\sigma ^{\prime }}^{R}$
and $\mathcal{G}_{i\sigma ,j\sigma ^{\prime }}^{A}$ . The
matrices $\Gamma ^{(l)}$ are given by the self-energy due to contact $l$, $%
\Gamma ^{(l)}\!=\!\mathrm{i}[\Sigma _{l}^{R}\!-\!\Sigma _{l}^{A}]$ where $%
\Sigma _{l}^{R}\!$ and $\!\Sigma _{l}^{A}$ are the self-energies matrices of the
retarded and advanced propagators respectively. Note that the definition of $T$
includes the spin index.

A typical $T$ vs. $B_{z}$ signal for strong spin-orbit coupling is shown in
Fig.\ref{Figu1}(a). A splitting of the first focusing peak is clearly observed
\cite{UsajB04_focusing}. Notably, there is no splitting of the second peak.
These results can be easily interpreted in terms of the semiclassical picture
given above. From all the semiclassical orbits that connect the $I$ and $D$
contacts, the ones that give the largest contribution to $T$ are those with
$2r_{\pm }\!=L$ \cite{UsajB04_focusing,vanHouten89}. When the applied magnetic field $%
B_{z}$ is increased the cyclotron radii are reduced as $B_{z}^{-1}$and the first
maximum in the transmission is found when $r_{-}(B_{z})\!=\!L/2$ as
schematically shown in Fig.\ref{Figu1}.(b). There $O_{1}$ is the electron path
between $I$ and $D$, this path is a semicircle of radius $r_{-}$. For this
field, indicated as $B_{z}=B_{1,1}$, the electrons that flow out of the injector
in the $O_{2}$ orbit do not
arrive to the detector since $r_{+}(B_{1,1})>L/2$. Furthermore, the two orbits $%
O_{1}$ and $O_{2}$ correspond to electrons injected with spin $down$ or $up$ in
the $y$-direction, respectively. Note that due to the SO coupling, the spin
rotates along the orbit. It is convenient to split the total transmission in the
four contributions $T_{\alpha \beta }$ corresponding to electrons injected with
spin $\alpha $ and collected with spin $\beta $. The total transmission can be
put as $T=T_{uu}+T_{ud}+T_{du}+T_{dd}$ and for $B_{z}=B_{1,1}$ the total
transmittance is dominated by the contribution $T_{du}$. When $B_{z}$ is
increased over $B_{1,1}$, $r_{-}(B_{1,1})<L/2$ and $T$ decreases. The next
maximum is reached for $B_{z}\!=\!B_{1,2}$ when $r_{+}(B_{1,2})\!=\!L/2$ and the
relevant orbit is $O_{2}$ as shown in Fig.\ref{Figu1}.(c). For this focusing
field the transmission is dominated by $T_{ud}$.

The next maximum in $T$ is found when $B_{z}\!=\!B_{2}$ and corresponds to the
situation shown in Fig.\ref{Figu1}.(d). This focusing condition is due to the
semiclassical trajectories with one intermediate bounce at the edge of the
sample. In this case the two possible paths $O_{1}$ and $O_{2}$ contribute to
the transmission. Electrons leaving the injector with a given spin arrive at the
detector with the same spin projection. Accordingly, the total coefficient $T$
is dominated by $T_{uu}\!+\!T_{dd}$. Clearly, $B_{2}$ is the magnetic field for
which $2(r_{-}\!+\!r_{+})\!=\!L$ holds. In agreement with the exact numerical
result, by extrapolating the semiclassical picture shown in Fig.\ref{Figu1}, one
finds that the peaks that are split due to Rashba interaction are those in which
the number of bounces is even (or zero).

\section{Imaging Techniques in Transverse Focusing with spin-orbit coupling}

\begin{figure}[t]
  \begin{center}
     \includegraphics[width=.8\textwidth   ]{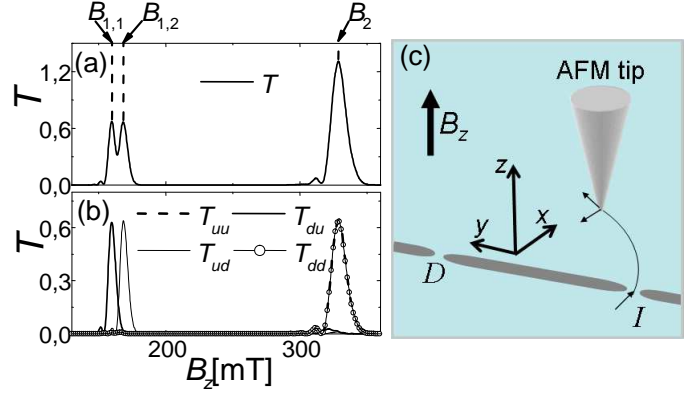}
     \caption{(a) Total focusing transmission coefficient $T$ versus applied perpendicular magnetic field $B_z$. (b) Spin resolved transmission coefficients versus
     $B_z$. We used $E_F\! = \! 23$meV, $m^{*}\! = \!0.055m_0$, $\alpha\! = \! 7$meVnm,
     $\beta\! = \!0$, $L\! = \!1.5\mu m$ and the width of the contacts is $70 nm$. (c) Schematic of an SPM imaging procedure.}
   \label{Figu2}
  \end{center}
\end{figure}

Scanning probe microscopy (SPM) techniques have been recently used for imaging
the electron flow in a variety of 2DEG ballistic systems
\cite{TopinkaLSHWMG00,TopinkaLWSFHMG01}. With this technique, the negatively
charged tip of a scanning microscope is positioned above the 2DEG as
schematically shown in Fig.\ref{Figu2}(c). The tip position can be changed to
sweep a given area of the explored 2D device. The electrons under the tip are
repelled and consequently a zone of lower electron density (or $divot$) is
formed under the tip. In the simplest case the transmission (and then the
conductance) between two contacts of the device is measured as the tip position
changes. If the tip is located in a region that affects the electron path
between the contacts the conductance changes providing a map of the electron
flow in the device. The resolution of these images is smaller than the divot
size, \cite{TopinkaLSHWMG00,TopinkaLWSFHMG01} making this technique a powerful
tool for studying nano-scale ballistic systems.

Here, we propose the use of this technique to explore the transverse focusing in
the presence of spin-orbit interaction \cite{note2}. We simulate the effect of
the tip potential by perturbing (increasing) the site energies $\varepsilon
_{i,\sigma }$ in an area of the order of $100nm^{2}$ centered at the tip
position. The Dyson equation is used to introduce the perturbation and the exact
propagators between the contacts $I$ and $D$ are calculated for each position of
the tip.

\begin{figure}[t]
  \begin{center}
     \includegraphics[width=.6\textwidth]{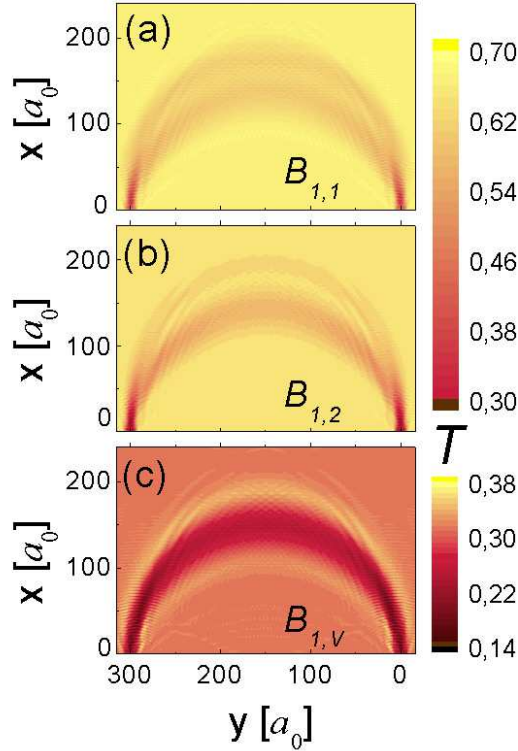}
     \caption{Total transmission coefficient $T$ between the contacts 1 and 2 as a function of the probe position for (a) $B_z\! = \! B_{1,1}$ , (b) $B_z\! = \! B_{1,2}$ and (c) $B_z\! = \! B_{1,v}\! = \! (B_{1,1}+B_{1,2})/2$. We used $\alpha \! = \! 7$meVnm and the parameters given in Fig.\ref{Figu2}. }
   \label{Figu3}
  \end{center}
\end{figure}

Figure \ref{Figu3}.(a) shows $T$ vs the tip's position when the perpendicular
magnetic field is fixed to obtain the first maximum ($B_{z}\!=\!B_{1,1}$) for a
SO coupling $\alpha \!=\!7$meVnm. The semicircular electron path is clearly
observed. In this case the $T$ map is dominated by a drop in $T$ along the
$O_{1}$ path. A similar pattern is found for the second transmission maximum
($B_{z}\!=\!B_{1,2}$) as shown in Fig.\ref{Figu3}.(b). In this case, the drop in
$T$ is due to the scattering induced by the tip of electrons travelling along
the $O_{2}$ path. A slightly different situation is found when $B_{z}$ is fixed
in between $B_{1,1}$ and $B_{1,2}$ as shown in Fig.\ref{Figu3}.(c); although the
variation is also dominated by a drop (dark area), $T$ increases at the two
sides of the minimum. The observation of these two lobes shows that the tip,
when placed at those positions, modifies the electron flow making a
non-focalized electron path---$O_{1}$ or $O_{2}$ in Fig.\ref{Figu1}.(b)-(c)---to
contribute to the transmission.

\begin{figure}[t]
  \begin{center}
     \includegraphics[width=.8\textwidth]{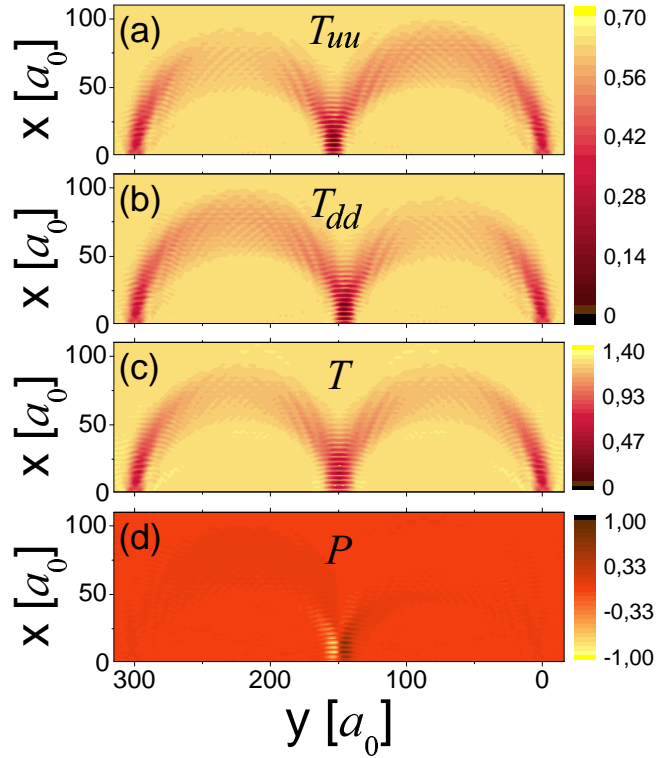}
     \caption{ We plot (a) $T_{uu}$, (b) $T_{dd}$, (c) $T$ and (d) $P$ as a function of the probe position for $B_z\! = \! B_2$. We used $\alpha \! = \! 7$meVnm and the parameters given in Fig.\ref{Figu2}.}
   \label{Figu4}
  \end{center}
\end{figure}

\begin{figure}[t]
  \begin{center}
     \includegraphics[width=.8\textwidth]{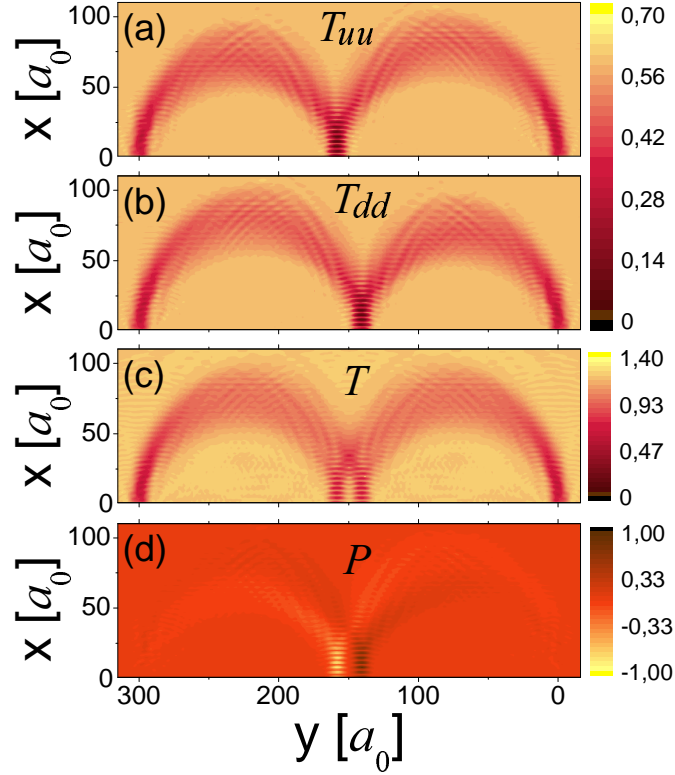}
     \caption{ We plot (a) $T_{uu}$, (b) $T_{dd}$, (c) $T$ and (d) $P$ as a function of the probe position for $B_z\! = \! B_2$. We used $\alpha \! = \! 15$meVnm and the parameters given in Fig.\ref{Figu2}.}
   \label{Figu5}
  \end{center}
\end{figure}

More interesting are the imaging results obtained when the field is fixed at the
second focusing condition: $B_{z}\!=\!B_{2}$. As mentioned above, in this case
$T$ is dominated by the electron's orbits with one bounce at the sample's edge.
For this field the largest contributions to the transmission coefficient are
$T_{uu}$ and $T_{dd}$ , and the corresponding focusing peak is unsplit. In
Fig.\ref{Figu4} and Fig.\ref
{Figu5} the results for this case are shown for $\alpha \!=\!7$meVnm and $%
15$meVnm, respectively. Panel (a) shows $T_{uu}$ as a function of the position
of the microscope probe. The change in the transmission in this case
clearly shows that the electrons injected with spin $up$ (in the $y$%
-direction) leave the injector in the bigger orbit, rebound and then arrive to
the detector in the smaller orbit with spin $up$---see $O_{2}$ in Fig.\ref
{Figu1}(d). In panel (b) the transmittance $T_{dd}$ is shown ---see $O_{1}$ in
Fig.\ref{Figu1}(d) - and in panel (c) the total transmission coefficient is
presented. As $T_{ud}$ and $T_{du}$ are very small, the total transmission is
essentially given by the sum of the contributions shown in panels (a) and (b).
Experimentally these two contributions could be distinguished by selecting the
spin of the injected carriers. In fact, a combination of an external in-plane
magnetic field in the $y$-direction and an appropriate gate voltage in the point
contacts can be used to filter spins in the injector or detector
\cite{PotokFMU02}. Selecting the spin of the injected electrons would make it
possible to separate the two trajectories---(a) and (b) in Figs. \ref{Figu4} and
\ref{Figu5}---and obtain a direct visualization of the two orbits split by the
spin-orbit coupling. Conversely, selecting the spin in the detector D, the
transmissions $T_{+}=T_{uu}+T_{du}$ and $T_{-}=T_{ud}+T_{dd}$ of carriers
arriving at $D$ with spin $up$ and $down$, respectively, could be measured. In
terms of these quantities, we define the polarization $P$ of the transmitted
particles as
\[
P=\frac{T_{+}-T_{-}}{T}
\]

Panel (c) and (d) of Figs. \ref{Figu4} and \ref{Figu5} show the total
transmission coefficient $T\!$ \ and the polarization $P$ as a function of the
tip position. The two semicircular electron paths including the rebound at the
edge are visualized in the $T$ map. In our simulations the smaller and the
bigger electron paths are not easily resolved in total transmission coefficient
map except for the largest SO coupling case and for the tip close to the bounce
position---see Fig.\ref{Figu5}(c). There, an appreciable fall (about $50\%$) of
the transmission in the two rebound positions indicates that, when the probe is
positioned there, the contribution to $T$ of one of the two possible electron
paths ($O_{1}$ or $O_{2}$) is being suppressed. If $O_{1}$ is being suppressed,
the electrons arriving to the detector will have spin $up$. On the other hand,
if $O_{2}$ is suppressed only spin $down$ electrons will arrive to the detector.
This means that one can select the spin polarization of the outgoing carrier
flux by changing the tip position a few nanometers as shown in
Fig.\ref{Figu4}(d) and Fig.\ref{Figu5}(d). Notably, the effect is also clearly
observed in the case of the smaller SO coupling despite of the fact that the
total transmittance $T$ does not resolve the two orbits.

\section{Summary and Conclusions}

We have discussed a microscopy imaging technique for the case transverse
electron focusing in 2DEGs with strong Rashba coupling. The main results can be
summarized as follows:

\textit{i.- }The existence of two different cyclotron radii splits the first
focusing peak onto two sub-peaks, each one corresponds to electrons arriving to
the detector with different spin polarization along the direction parallel to
the sample's edge.

\textit{ii.- }The images of the electron flow for focusing fields corresponding
to the first two sub-peaks are very similar and consequently, for this case, the
technique can not clearly distinguish the two type of orbits.

\textit{iii.- }When the external field is fixed between the focusing fields of
the two sub-peaks, $B_{z}=(B_{1,1}+B_{1,2})/2$, the transmission map shows a
structure that indicates the presence of the two orbits.

\textit{iv.-} For the second focusing condition, and for the case of strong
Rashba coupling, the technique can resolve the two orbits when the microscope
tip is placed close to the rebound position.

\textit{v.- }For the case described in the previous point, the microscope tip
can be used to polarize the electron flux arriving at the detector. The
direction of the polarization can be reversed by changing the tip position a few
nanometers.

Finally, we would like to emphasize a few points: \textit{a.-} Interference
fringes, characteristic of the quantum ballistic transport regime, are observed
in all the $T$ maps; \textit{b.-} For the properties studied here, replacing the
hard wall potential $V(x)$ by a more realistic parabolic potential does not
change the main properties of the system \cite{UsajB05_SHE,GovorovKD04}.
Therefore, our results should correctly describe the images that could be
obtained in heterostructures defined by gates. \textit{c.- } The competition
between the Rashba and the Dresselhauss couplings leads to interesting features
in the focusing signal and needs to be considered for interpreting imaging
results in systems where these two SO interactions are present. These results
will be presented elsewhere.

\section{Acknowledgment}
This work was supported by ANPCyT Grants No 13829 and 13476 and CONICET PIP
5254. AR acknowledge support from PITP and CONICET. GU is a member of CONICET.


\begin{thebibliography}{99.}
\expandafter\ifx\csname natexlab\endcsname\relax\def\natexlab#1{#1}\fi
\expandafter\ifx\csname bibnamefont\endcsname\relax
  \def\bibnamefont#1{#1}\fi
\expandafter\ifx\csname bibfnamefont\endcsname\relax
  \def\bibfnamefont#1{#1}\fi
\expandafter\ifx\csname citenamefont\endcsname\relax
  \def\citenamefont#1{#1}\fi
\expandafter\ifx\csname url\endcsname\relax
  \def\url#1{\texttt{#1}}\fi
\expandafter\ifx\csname urlprefix\endcsname\relax\def\urlprefix{URL }\fi
\providecommand{\bibinfo}[2]{#2} \providecommand{\eprint}[2][]{\url{#2}}

\bibitem{Spintronicsbook}
\bibinfo{editor}{\bibfnamefont{D.}~\bibnamefont{Awschalom}},
  \bibinfo{editor}{\bibfnamefont{N.}~\bibnamefont{Samarth}}, \bibnamefont{and}
  \bibinfo{editor}{\bibfnamefont{D.}~\bibnamefont{Loss}}, eds.,
  \emph{\bibinfo{title}{Semiconductor Spintronics and Quantum Computation}}
  (\bibinfo{publisher}{Springer}, \bibinfo{address}{New York},
  \bibinfo{year}{2002}).

\bibitem{ReynosoUB06}
\bibinfo{author}{\bibfnamefont{A.}~\bibnamefont{Reynoso}},
  \bibinfo{author}{\bibfnamefont{G.}~\bibnamefont{Usaj}}, \bibnamefont{and}
  \bibinfo{author}{\bibfnamefont{C.~A.} \bibnamefont{Balseiro}},
  \bibinfo{journal}{Phys. Rev. B} \textbf{\bibinfo{volume}{73}},
  \bibinfo{pages}{115342} (\bibinfo{year}{2006}).

\bibitem{Winkler_book}
\bibinfo{author}{\bibfnamefont{R.}~\bibnamefont{Winkler}},
  \emph{\bibinfo{title}{Spin-orbit coupling effects in two-dimensional electron
  and hole systems}} (\bibinfo{publisher}{Springer-Verlag},
  \bibinfo{year}{2003}).

\bibitem{NittaATE97}
\bibinfo{author}{\bibfnamefont{J.}~\bibnamefont{Nitta}},
  \bibinfo{author}{\bibfnamefont{T.}~\bibnamefont{Akazaki}},
  \bibinfo{author}{\bibfnamefont{H.}~\bibnamefont{Takayanagi}},
  \bibnamefont{and} \bibinfo{author}{\bibfnamefont{T.}~\bibnamefont{Enoki}},
  \bibinfo{journal}{Phys. Rev. Lett.} \textbf{\bibinfo{volume}{78}},
  \bibinfo{pages}{1335} (\bibinfo{year}{1997}).

\bibitem{TopinkaLSHWMG00}
\bibinfo{author}{\bibfnamefont{M.}~\bibnamefont{Topinka}},
  \bibinfo{author}{\bibfnamefont{B.}~\bibnamefont{LeRoy}},
  \bibinfo{author}{\bibfnamefont{S.}~\bibnamefont{Shaw}},
  \bibinfo{author}{\bibfnamefont{E.}~\bibnamefont{Heller}},
  \bibinfo{author}{\bibfnamefont{R.}~\bibnamefont{Westervelt}},
  \bibinfo{author}{\bibfnamefont{K.}~\bibnamefont{Maranowski}},
  \bibnamefont{and} \bibinfo{author}{\bibfnamefont{A.}~\bibnamefont{Gossard}},
  \bibinfo{journal}{Science}  (\bibinfo{year}{2000}).

\bibitem{TopinkaLWSFHMG01}
\bibinfo{author}{\bibfnamefont{M.}~\bibnamefont{Topinka}},
  \bibinfo{author}{\bibfnamefont{B.}~\bibnamefont{LeRoy}},
  \bibinfo{author}{\bibfnamefont{R.}~\bibnamefont{Westervelt}},
  \bibinfo{author}{\bibfnamefont{S.}~\bibnamefont{Shaw}},
  \bibinfo{author}{\bibnamefont{R.Fleischmann}},
  \bibinfo{author}{\bibfnamefont{E.}~\bibnamefont{Heller}},
  \bibinfo{author}{\bibfnamefont{K.}~\bibnamefont{Maranowski}},
  \bibnamefont{and} \bibinfo{author}{\bibfnamefont{A.}~\bibnamefont{Gossard}},
  \bibinfo{journal}{Nature} \textbf{\bibinfo{volume}{410}},
  \bibinfo{pages}{183} (\bibinfo{year}{2001}).

\bibitem{UsajB04_focusing}
\bibinfo{author}{\bibfnamefont{G.}~\bibnamefont{Usaj}} \bibnamefont{and}
  \bibinfo{author}{\bibfnamefont{C.~A.} \bibnamefont{Balseiro}},
  \bibinfo{journal}{Phys. Rev. B} \textbf{\bibinfo{volume}{70}},
  \bibinfo{pages}{041301(R)} (\bibinfo{year}{2004}).

\bibitem{BychkovR84}
\bibinfo{author}{\bibfnamefont{Y.~A.} \bibnamefont{Bychkov}} \bibnamefont{and}
  \bibinfo{author}{\bibfnamefont{E.~I.} \bibnamefont{Rashba}},
  \bibinfo{journal}{JETP Letters} \textbf{\bibinfo{volume}{39}},
  \bibinfo{pages}{78} (\bibinfo{year}{1984}).

\bibitem{note1}
\bibinfo{note}{These are the solutions for positive $B_{z}$. For negative
  $B_{z}$ the eingenstates change: $\Psi _{n,k}^{\pm
  }|_{-|B_{z}|}=\mathrm{i}\sigma _{y}\Psi _{n,k}^{\pm }|_{|B_{z}|}$.}

\bibitem{ReynosoUSB04}
\bibinfo{author}{\bibfnamefont{A.}~\bibnamefont{Reynoso}},
  \bibinfo{author}{\bibfnamefont{G.}~\bibnamefont{Usaj}},
  \bibinfo{author}{\bibfnamefont{M.~J.} \bibnamefont{Sanchez}},
  \bibnamefont{and} \bibinfo{author}{\bibfnamefont{C.~A.}
  \bibnamefont{Balseiro}}, \bibinfo{journal}{Phys. Rev. B}
  \textbf{\bibinfo{volume}{70}}, \bibinfo{pages}{235344}
  (\bibinfo{year}{2004}).

\bibitem{PletyukhovAMB02}
\bibinfo{author}{\bibfnamefont{M.}~\bibnamefont{Pletyukhov}},
  \bibinfo{author}{\bibfnamefont{C.}~\bibnamefont{Amann}},
  \bibinfo{author}{\bibfnamefont{M.}~\bibnamefont{Mehta}}, \bibnamefont{and}
  \bibinfo{author}{\bibfnamefont{M.}~\bibnamefont{Brack}},
  \bibinfo{journal}{Phys. Rev. Lett.} \textbf{\bibinfo{volume}{89}},
  \bibinfo{pages}{116601} (\bibinfo{year}{2002}).

\bibitem{GutzBook}
\bibinfo{author}{\bibfnamefont{M.}~\bibnamefont{Gutzwiller}},
  \emph{\bibinfo{title}{Chaos in Classical and Quantum Mechanics}}
  (\bibinfo{publisher}{Spring-Verlag}, \bibinfo{address}{New York},
  \bibinfo{year}{1991}).

\bibitem{PotokFMU02}
\bibinfo{author}{\bibfnamefont{R.~M.} \bibnamefont{Potok}},
  \bibinfo{author}{\bibfnamefont{J.~A.} \bibnamefont{Folk}},
  \bibinfo{author}{\bibfnamefont{C.~M.} \bibnamefont{Marcus}},
  \bibnamefont{and} \bibinfo{author}{\bibfnamefont{V.}~\bibnamefont{Umansky}},
  \bibinfo{journal}{Phys. Rev. Lett.} \textbf{\bibinfo{volume}{89}},
  \bibinfo{pages}{266602} (\bibinfo{year}{2002}).

\bibitem{RokhinsonLGPW04}
\bibinfo{author}{\bibfnamefont{L.~P.} \bibnamefont{Rokhinson}},
  \bibinfo{author}{\bibfnamefont{V.}~\bibnamefont{Larkina}},
  \bibinfo{author}{\bibfnamefont{Y.~B.} \bibnamefont{Lyanda-Geller}},
  \bibinfo{author}{\bibfnamefont{L.~N.} \bibnamefont{Pfeiffer}},
  \bibnamefont{and} \bibinfo{author}{\bibfnamefont{K.~W.} \bibnamefont{West}},
  \bibinfo{journal}{Phys. Rev. Lett.} \textbf{\bibinfo{volume}{93}},
  \bibinfo{pages}{146601} (\bibinfo{year}{2004}).

\bibitem{vanHouten89}
\bibinfo{author}{\bibfnamefont{H.}~\bibnamefont{van Houten}},
  \bibinfo{author}{\bibfnamefont{C.~W.~J.} \bibnamefont{Beenakker}},
  \bibinfo{author}{\bibfnamefont{J.~G.} \bibnamefont{Willianson}},
  \bibinfo{author}{\bibfnamefont{M.~E.~I.} \bibnamefont{Broekaart}},
  \bibinfo{author}{\bibfnamefont{P.~H.~M.} \bibnamefont{Loosdrecht}},
  \bibinfo{author}{\bibfnamefont{B.~J.} \bibnamefont{van Wees}},
  \bibinfo{author}{\bibfnamefont{J.~E.} \bibnamefont{Mooji}},
  \bibinfo{author}{\bibfnamefont{C.~T.} \bibnamefont{Foxon}}, \bibnamefont{and}
  \bibinfo{author}{\bibfnamefont{J.~J.} \bibnamefont{Harris}},
  \bibinfo{journal}{Phys. Rev. B} \textbf{\bibinfo{volume}{39}},
  \bibinfo{pages}{8556} (\bibinfo{year}{1989}).

\bibitem{BeenakkerH91}
\bibinfo{author}{\bibfnamefont{C.~W.} \bibnamefont{Beenakker}}
  \bibnamefont{and} \bibinfo{author}{\bibfnamefont{H.}~\bibnamefont{van
  Houten}}, in \emph{\bibinfo{booktitle}{Solid State Physics}}, edited by
  \bibinfo{editor}{\bibfnamefont{H.}~\bibnamefont{Eherenreich}}
  \bibnamefont{and} \bibinfo{editor}{\bibfnamefont{D.}~\bibnamefont{Turnbull}}
  (\bibinfo{publisher}{Academic Press}, \bibinfo{address}{Boston},
  \bibinfo{year}{1991}), vol.~\bibinfo{volume}{44}, pp.
  \bibinfo{pages}{1--228}.

\bibitem{Ferrybook}
\bibinfo{author}{\bibfnamefont{D.~K.} \bibnamefont{Ferry}} \bibnamefont{and}
  \bibinfo{author}{\bibfnamefont{S.~M.} \bibnamefont{Goodnick}},
  \emph{\bibinfo{title}{Transport in Nanostructures}}
  (\bibinfo{publisher}{Cambridge University Press}, \bibinfo{address}{New
  York}, \bibinfo{year}{1997}).

\bibitem{note2}
\bibinfo{note}{Imaging of cyclotron orbits using this imaging technique was
  reported at ICPS 28, Vienna (2006) by K. Aidala. See also \cite{AidalaPHW06,AidalaPHW07}}.

\bibitem{UsajB05_SHE}
\bibinfo{author}{\bibfnamefont{G.}~\bibnamefont{Usaj}} \bibnamefont{and}
  \bibinfo{author}{\bibfnamefont{C.~A.} \bibnamefont{Balseiro}},
  \bibinfo{journal}{Europhys. Lett.} \textbf{\bibinfo{volume}{72}},
  \bibinfo{pages}{631} (\bibinfo{year}{2005}).

\bibitem{GovorovKD04}
\bibinfo{author}{\bibfnamefont{A.~O.} \bibnamefont{Govorov}},
  \bibinfo{author}{\bibfnamefont{A.~V.} \bibnamefont{Kalameitsev}},
  \bibnamefont{and} \bibinfo{author}{\bibfnamefont{J.~P.} \bibnamefont{Dulka}},
  \bibinfo{journal}{Phys. Rev. B} \textbf{\bibinfo{volume}{70}},
  \bibinfo{pages}{245310} (\bibinfo{year}{2004}).


\bibitem{SchliemannL03}
\bibinfo{author}{\bibfnamefont{J.}~\bibnamefont{Schliemann}} \bibnamefont{and}
  \bibinfo{author}{\bibfnamefont{D.}~\bibnamefont{Loss}},
  \bibinfo{journal}{Phys. Rev. B} \textbf{\bibinfo{volume}{68}},
  \bibinfo{pages}{165311} (\bibinfo{year}{2003}).

\bibitem{AidalaPHW06}
\bibinfo{author}{\bibfnamefont{K.~E.} \bibnamefont{Aidala}},
  \bibinfo{author}{\bibfnamefont{R.~E.} \bibnamefont{Parrott}},
  \bibinfo{author}{\bibfnamefont{E.}~\bibnamefont{Heller}}, \bibnamefont{and}
  \bibinfo{author}{\bibfnamefont{R.}~\bibnamefont{Westervelt}},
  \bibinfo{journal}{Int. J. Mod. Phys. A}
  \textbf{\bibinfo{volume}{21}},  \bibinfo{pages}{4407-4424} (\bibinfo{year}{2006}).



\bibitem{AidalaPHW07}
\bibinfo{author}{\bibfnamefont{K.~E.} \bibnamefont{Aidala}},
  \bibinfo{author}{\bibfnamefont{R.~E.} \bibnamefont{Parrott}},
    \bibinfo{author}{\bibfnamefont{Tobias} \bibnamefont{Kramer}},
  \bibinfo{author}{\bibfnamefont{E.~J.}~\bibnamefont{Heller}},
  \bibinfo{author}{\bibfnamefont{R.~M.}~\bibnamefont{Westervelt}},
  \bibinfo{author}{\bibfnamefont{M.~P.}~\bibnamefont{Hanson}},
 \bibnamefont{and}
  \bibinfo{author}{\bibfnamefont{A.~C.}~\bibnamefont{Gossard}},
    \bibinfo{journal}{Nat. Phys.} \textbf{\bibinfo{volume}{3}},  \bibinfo{pages}{464-468}
  (\bibinfo{year}{2007}).

\end{thebibliography}
\end{document}